\documentclass[aps,prb,superscriptaddress,twocolumn,showpacs,amsmath,floatfix,citeautoscript]{revtex4}
\usepackage{graphicx}
\usepackage{float}
\usepackage{dcolumn}
\usepackage{color}
\usepackage{latexsym,bm}
\usepackage[normalem]{ulem}
\usepackage{multirow}
\usepackage{appendix}
\usepackage{epstopdf}

\begin{document}

\hyphenpenalty=5000
\tolerance=1000

\title{Magnetic ground state and electron doping tuning of Curie temperature
in Fe$_3$GeTe$_2$: first-principles studies}
\author{Zhen-Xiong Shen}
\affiliation{Key Laboratory of Quantum Information, University of Science and
  Technology of China, Hefei, Anhui, 230026, People's Republic of China}
\affiliation{Synergetic Innovation Center of Quantum Information and Quantum
  Physics, University of Science and Technology of China, Hefei, 230026, People's Republic of China}
\author{Xiangyan Bo}
\affiliation{National Laboratory of Solid State Microstructures and School of Physics,
Nanjing University, Nanjing 210093, China}
\affiliation{Collaborative Innovation Center of Advanced Microstructures, Nanjing
University, Nanjing 210093, China}
\author{Kun Cao}
\affiliation{Daresbury Laboratory, Daresbury, Warrington WA4 4AD, United Kingdom}
\author{Xiangang Wan}
\email{xgwan@nju.edu.cn}
\affiliation{National Laboratory of Solid State Microstructures and School of Physics,
Nanjing University, Nanjing 210093, China}
\affiliation{Collaborative Innovation Center of Advanced Microstructures, Nanjing
University, Nanjing 210093, China}
\author{Lixin He}
\email{helx@ustc.edu.cn}
\affiliation{Key Laboratory of Quantum Information, University of Science and
  Technology of China, Hefei, Anhui, 230026,  People's Republic of China}
\affiliation{Synergetic Innovation Center of Quantum Information and Quantum
  Physics, University of Science and Technology of China, Hefei, 230026, People's Republic of China}
\date{\today}

\begin{abstract}
Intrinsic magnetic van der Waals (vdW) materials have attracted much attention, especially Fe$_{3}$GeTe$_{2}$ (FGT), which exhibits highly tunable properties. However, despite vast efforts, there are still several challenging issues to be resolved. Here using a first-principles linear-response approach, we carry out comprehensive investigation of both bulk and monolayer FGT. We find that although the magnetic exchange interactions in FGT are frustrated, our Monte Carlo simulations agree with the total energy calculations, confirming that the ground state of bulk FGT is indeed ferromagnetic (FM). A tiny electron doping reduces the magnetic frustration, resulting in significant increasing of the Curie temperature. We also calculate the spin-wave dispersion, and find a small spin gap as well as a nearly flat band in the magnon spectra. These features can be used to compare with the future neutron scattering measurement and finally clarify the microscopic magnetic mechanism in this two-dimensional family materials.
\end{abstract}

\maketitle

{\it Introduction:}
The searching for two-dimensional (2D) materials with novel properties has been driven by
continuous development of modern devices applications.
Recently, 2D ferromagnetic (FM) materials, such as, Cr$_2$Ge$_2$Te$_6$  \cite{gong2017discovery},
CrI$_3$ \cite{huang2017layer-dependent}, and Fe$_3$GeTe$_2$ (FGT) \cite{deng2018gate}
have been fabricated, which have attracted great attention, because of their potential
applications in spintronic devices, as well as fundamental interests in low dimensional magnetisms.
Among the known 2D FM materials, FGT is of special interests \cite{deng2018gate,fei2018two,kim2018large,gibertini2019magnetic,johansen2019current}, because it is
the only metallic 2D FM material synthesized so far. Moreover, the thin layer FGT displays novel gate-controlled properties \cite{deng2018gate}. Its rich physics due to the coupling between the charge and spin degree of freedom as well as the potential applications as FM electrodes  in spintronic devices attracted a lot of research attentions \cite{deng2018gate,fei2018two,kim2018large,gibertini2019magnetic,johansen2019current,
ding2020observation,fang2019observation,xu2019large,verchenko2015ferromagnetic,park2020controlling,zhu2016electronic,
zhuang2016strong,liu2017critical,liu2017wafer,yi2016competing,tan2018hard,zhang2018emergence,albarakati2019antisymmetric,alghamdi2019highly}.

Bulk FGT is a layered van der Waals crystal, and the ground state of the system had been proposed as FM by various experiments\cite{chen2013magnetic,deng2018gate,liu2017wafer,fei2018two,wang2019current-driven,verchenko2015ferromagnetic}.
A high out-of-plane magnetocrystalline anisotropy had been suggested\cite{verchenko2015ferromagnetic,leon2016magnetic}. Very recently, experiments demonstrated that FGT may retain FM order even down to the
atomically thin monolayer with a relatively high transition temperature, about 130 K \cite{fei2018two}.  Remarkably, the transition temperature of a tri-layer FGT can be elevated to room temperature
by gate-controlled electron doping\cite{deng2018gate}. However, there are also experimental and theoretical works \cite{yi2016competing,C9NR10171C}
emphasize the competing and coexisting FM and anti-ferromagnetic (AFM) states in this materials, and the role of Fe-defect and electron doping had also been discussed \cite{C9NR10171C}. To clarify the intrinsic magnetic order in the stoichiometric FGT is an important issue. Furthermore, usually an small carrier doping cannot change the magnetic properties significantly. Therefore to understand the microscopic mechanism response to the strong enhancement of Curie temperature induced by an ionic gate is of great importance for the potential voltage-controlled magnetoelectronics based on atomically thin van der Waals crystals.

In this letter, we carry out comprehensive first-principles
studies of the magnetic properties for both bulk and monolayer FGTs, as well as the effects of the electron doping on monolayer FGTs.
The linear-response calculations reveal that the magnetic exchange interactions are geometrically frustrated
within each Fe plane, whereas the inter-plane magnetic exchange interaction are strongly FM.
There are also competing magnetic exchange interactions between the top and bottom monolayer in a FGT unit cell.
The Monte Carlo simulations based on the obtained magnetic exchange interactions predict a FM ground state for bulk FGT, which confirms the total energy calculations. The estimated high T$_{c}$s are in good agreement with the experiments.\cite{deng2018gate,fei2018two,verchenko2015ferromagnetic,liu2017wafer,chen2013magnetic,deiseroth2006fe3gete2}
Electron doping may reduce the frustration of the magnetic exchange interactions, resulting in significant enhancement of the Curie temperature, which might explain the gate-tunable room-temperature FM observed in recent experiment.~\cite{deng2018gate} The calculation details are given in the Supplement Materials (SM) \cite{SM}.


\begin{figure}
  \centering
  \includegraphics[width=0.4\textwidth]{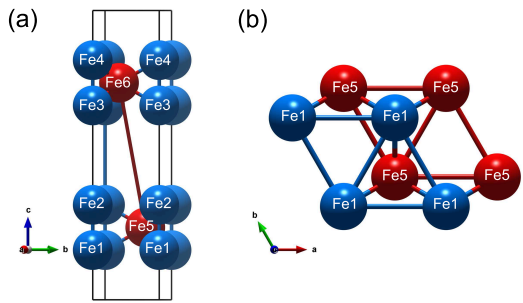}
  \caption{(Color online) (a) The Fe atoms in a unite cell of bulk FGT. There are two types of Fe ions. Type I Fe ions
   (Fe1 - Fe4) are in blue, whereas the Type II Fe ions (Fe5, Fe6) are in red.
   (b) Top view of the Fe atoms in a monolayer FGT. The Fe1 and Fe5 ions form triangular lattice respectively. }\label{fig:structure}
\end{figure}

{\it Bulk FGT:}
The bulk FGT has space group P6$_{3}$/mmc (No. 194). \cite{deiseroth2006fe3gete2,verchenko2015ferromagnetic}
The unit cell of FGT contains two monolayers, and in each monolayer, the Fe$_3$Ge heterometallic slab is sandwiched between two Te planes.
Figure~\ref{fig:structure} depicts the iron atoms in a FGT unit cell. There are six iron planes in a unit cell, and each plane contains an equivalent Fe ion, forming a triangular lattice. These Fe ions are labeled as Fe1 - Fe6, respectively. There are two types of Fe ions\cite{deiseroth2006fe3gete2}: Fe1 - Fe4 located at 4e Wyckoff position, while Fe5, Fe6 occupy 2c Wyckoff position.

Most experiments support that the ground state of FGT is FM\cite{chen2013magnetic,deng2018gate,liu2017wafer,fei2018two,wang2019current-driven,verchenko2015ferromagnetic}.
However, there are also some experiments and theories favoring of a AFM ground state\cite{yi2016competing,C9NR10171C}, i.e., within the top (down) monolayer, the Fe ions are FM, but between the top and down mono-layer, the spins are anti-parallel. To clarify the magnetic ground state of bulk FGT, we first compare the total energies of the FM and AFM state. Our results show that the ground state magnetic structure actually depend on the value of Coulomb $U$. The pure LSDA calculation, namely $U$=0, predicts a AFM order,
However, once a reasonable $U$ ($U >$ 2.5 eV) had been adopted, LSDA+U always suggests the FM state as more stable than AFM state (see Appendix for more details).
We note that in Refs. \onlinecite{yi2016competing,C9NR10171C}, the AFM ground were predicted by comparing the total energy of the FM and AFM states. However, the  Coulomb $U$ were ignored in their calculations.

While the magnetic moment from LSDA+U calculation for the FM spin ordering is in good agreements with the experiments, our calculations reveal that magnetic moments of Fe ions depend strongly on the magnetic configurations. For example, when we force the magnetic ground state to be a in-plane AFM order (i.e., FM along $c$-axis and AFM along $a$ and $b$-axis), the magnetic moments reduce from 1.6 $\mu_B$ of FM ordering to about 0.9  $\mu_B$. This indicates the itinerant nature of the magnetism.


Having shown that the ground state of bulk FGT is FM for different LSDA+U schemes with reasonable $U$ values, we now further confirm it by magnetic exchange interactions. Because different magnetic configurations show different magnetic moments, thus the calculated total energy differences between various magnetic ordering come not only from the magnetic exchange interaction energies but also from magnetization energies. Consequently, the traditional method to fit $J$s to the total energy differences between various magnetic configurations is not suitable here. Instead we evaluate the magnetic exchange interactions $J(\mathbf{q})$ by using a linear--response approach \cite{wan2006calculation}.

\begin{figure}
  \centering
  \includegraphics[width=0.4\textwidth]{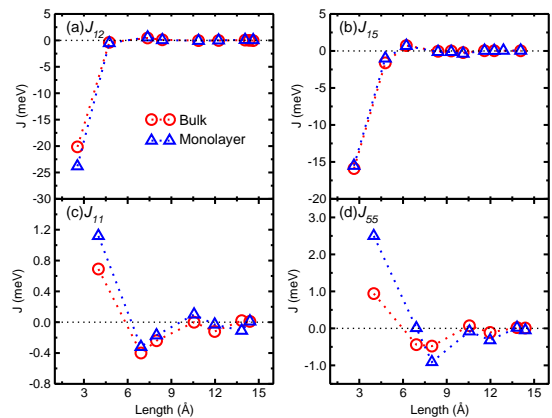}
  \caption{(Color online) The major exchange interactions (a) $J_{12}$, (b)$J_{15}$, (c)$J_{11}$, and (d) $J_{55}$
  in bulk (red circles) and monolayer (blue triangles) FGT as functions the distances between two spins. }\label{fig:exchangeinteraction}
\end{figure}

The exchange interactions are first calculated in the ${\bf q}$-space, and Fourier transferred to the real space.
We calculated all exchange interactions, including 12 sets $J_{i,j}(\Delta {\bf R})$ of intra-layer interactions (i.e., the $i$-th and $j$-th Fe ions belong to the same monolayer) and 9 sets of inter-layer interactions (the $i$-th and $j$-th Fe ions belong to different monolayers). Here, $\Delta {\bf R}$=${\bf R}_j$-${\bf R}_i$, whereas ${\bf R}_i$, and ${\bf R}_j$ are the unit cell vectors of the $i$-th and $j$-th spins. The selected exchange interactions as functions of distance between two spins are shown in Fig.~\ref{fig:exchangeinteraction}.
Most exchange interactions fall to small (but non-zero) values if the distances are larger than 10 \AA.
In Table S1 of the SM\cite{SM},
we list 23 largest exchange interactions.

The exchange interactions in FGT are rather complicated. We first consider the magnetic structures of the bottom monolayer, where the top monolayer has exactly the same magnetic structure and exchange interactions by symmetry. The bottom monolayer contains three planes of Fe ions, and in each plane the Fe ions
form a triangular lattice [see Fig.~\ref{fig:structure}(b)]. The exchange interactions between the nearest neighbor (NN) in-plane irons, $J_{1,1}$=0.69 meV, $J_{5,5}$= 0.94 meV are all AFM. This is a typical geometrically frustrated spin system, where the FM state is unstable if only the intra-plane interactions are considered.

The interactions between Fe1 and Fe2, $J_{1,2}$=-20.15 meV are strongly FM, which ensure that the spins of Fe1 (Fe3) and Fe2 (Fe4) ions aline parallel to each other. However, this does not guarantee the full lattice is FM, due to the intra-plane frustrated interactions ($J_{11}$, $J_{22}$ and $J_{55}$) discussed above. Crucially, the exchange interactions $J_{25}$=$J_{15}$=-15.89 meV are FM, which
force the spins of all three nearest Fe1, and Fe2 ions in the triangle
are all parallel to that of Fe5 ions, and thus stabilize the FM state.
As it will be shown in this work, that the frustrated spin interactions play crucial role in tuning the Curie temperatures.

Having understand the FM state in a monolayer, we now turn to the coupling between the top and bottom monolayers.
Interestingly, the NN inter-layer exchange interaction between two monolayers, $J_{2,3}$($\Delta {\bf R}$=0)=0.32 meV, which seems imply that the spins of the top and bottom  monolayers should have opposite directions, i.e., AFM.  However, the next-nearest spin interaction $J_{5,6}$($\Delta {\bf R}$=0)=-0.60 meV is FM and even stronger than $J_{2,3}$ against intuition. Furthermore, as shown in Table S1, the number of equivalent exchange interactions for $J_{2,3}$ and
$J_{5,6}$ are 2, 6 respectively. Therefore, the FM coupling between the two monolayers is stronger than the AFM coupling. Overall the ground state calculated from the Heisenberg model [see Eq.~(1) below] is FM instead of AFM, confirming the
direct total energy calculations.

\begin{figure}
  \centering
  \includegraphics[width=0.4 \textwidth]{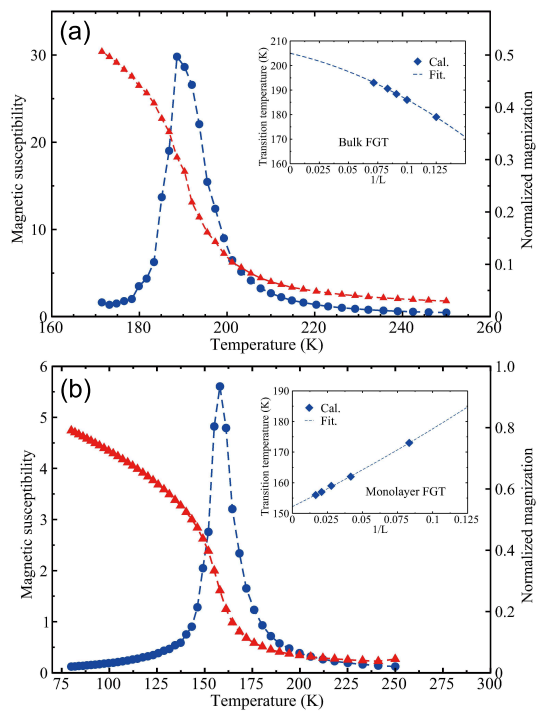}
    \caption{(Color online) The calculated susceptibility (blue dots)
    and magnetization (red triangles) as functions of temperature for (a) bulk FGT,
  simulated on a 12$\times$12$\times$12 lattice and (b) monolayer FGT,
  simulated on a 36$\times$36 lattice. The MAE is taken to be 0.34 meV/Fe ion.
  The inserts depict the transition temperatures $T_c$ as functions of $1/L$.
}\label{fig:tc}
\end{figure}

Although the magnetism is quite itinerant and the Stoner excitation may be strong here,
we still use the following Heisenberg model to estimate the magnetic transition temperature and the magnetic phase diagram \cite{deng2018gate}

\begin{equation}
H=\sum_{i<j} J_{ij} {\bf S}_i \cdot {\bf S}_j +\sum_i A(S_i^z)^2,
\end{equation}
where $A$ is the magnetic anisotropy energy (MAE). We adopt $L$$\times$$L$$\times$$L$ spuercells with $L$=8 - 14 in our re-MC simulations\cite{cao2009first}.
It is well known that the MAE depend strongly on the Coulomb $U$\cite{MAE-U}, and for this materials various calculations and experiments show that the MAE of FGT ranges from 0.34 --1.58 meV/Fe ion\cite{leon2016magnetic,verchenko2015ferromagnetic,zhuang2016strong,park2020controlling,deng2018gate}.
To avoid ambiguity we use the MAE as a parameter in the following simulations.
Figure~\ref{fig:tc}(a) depicts the magnetization and magnetic susceptibility of the $L$=12 lattice, where
the MAE is chosen to be 0.34 meV/Fe ion\cite{leon2016magnetic}.
The calculated FM Curie temperature for this lattice is around 191 K. The insert of
the figure shows the Curie temperature as a function of inverse lattice size $1/L$, where the transition temperature increases with increasing lattice sizes.
We perform finite size scaling to the Curie temperature and obtain $T_c$=205 K in the thermodynamic limit, which is in excellent agreements with the experimental values 200 - 230 K.
\cite{deng2018gate,fei2018two,verchenko2015ferromagnetic,liu2017wafer,chen2013magnetic,deiseroth2006fe3gete2}
We also exam how the MAE affects the transition temperature for the $L$=12 lattice.
As shown in Fig. S2 of SM\cite{SM}, the transition temperature increases a small mount
with the increasing of MAE as expected.

{\it Monolayer FGT:}
The monolayer FGT is phase stable\cite{zhuang2016strong, deng2018gate, fei2018two}, and can
maintain the long range FM order at reduced temperature $\sim$ 130 K
comparing to about 200 K in bulk FGT\cite{deiseroth2006fe3gete2, deng2018gate, fei2018two}.
We calculate the exchange interactions in the monolayer FGT
(i.e., half of the unit cell shown in Fig.~\ref{fig:structure}),
and the results are shown in Fig.~\ref{fig:exchangeinteraction} (blue triangles).
More details are listed in Table S2 in SM.
We find that the exchange interactions in monolayer FGT only change slightly compared with their counter parts in bulk FGT.
The first nearest neighbor exchange interaction value in monolayer FGT is $J_{1,2}$=-23.82 meV which is slightly stronger than that (-20.15 meV) in bulk FGT.  $J_{1,5}$ = -15.52 meV, which is approximately equal to -15.89 meV in bulk FGT. However, $J_{5,5}$=2.50 meV, which is about three times of that (0.94 meV) in bulk FGT. Also we have $J_{1,1}$=$J_{2,2}$=1.12 meV, which is about two times of 0.69 meV in bulk FGT.
Still $J_{1,5}$ is more than six times larger than $J_{1,1}$ and $J_{5,5}$ which ensures the FM ground state in the monolayer FGT.

We simulate the magnetic phase diagram of the monolayer FGT.
Figure~\ref{fig:tc}(b) depicts the magnetization and magnetic susceptibility as functions of temperature for a 36$\times$36 lattice.
Same as the bulk case, the MAE is set to 0.34 meV/Fe ion. The Curie temperature for these parameters are around 159 K. Interestingly, as shown in the insert of the figure, the Curie temperature deceases with the increasing size of lattice, in contrast to bulk cases. In the thermodynamic limit, the Curie temperature is
152 K, which is in close agreement with the experimental values 130 K.\cite{fei2018two}
We also investigate how the MAE change the Curie temperature. As shown in Fig.S2 of SM \cite{SM}, the Curie temperature also increase only slightly with the increasing MAE in monolayer FGT.

\begin{figure}
  \centering
  \includegraphics[width=0.4\textwidth]{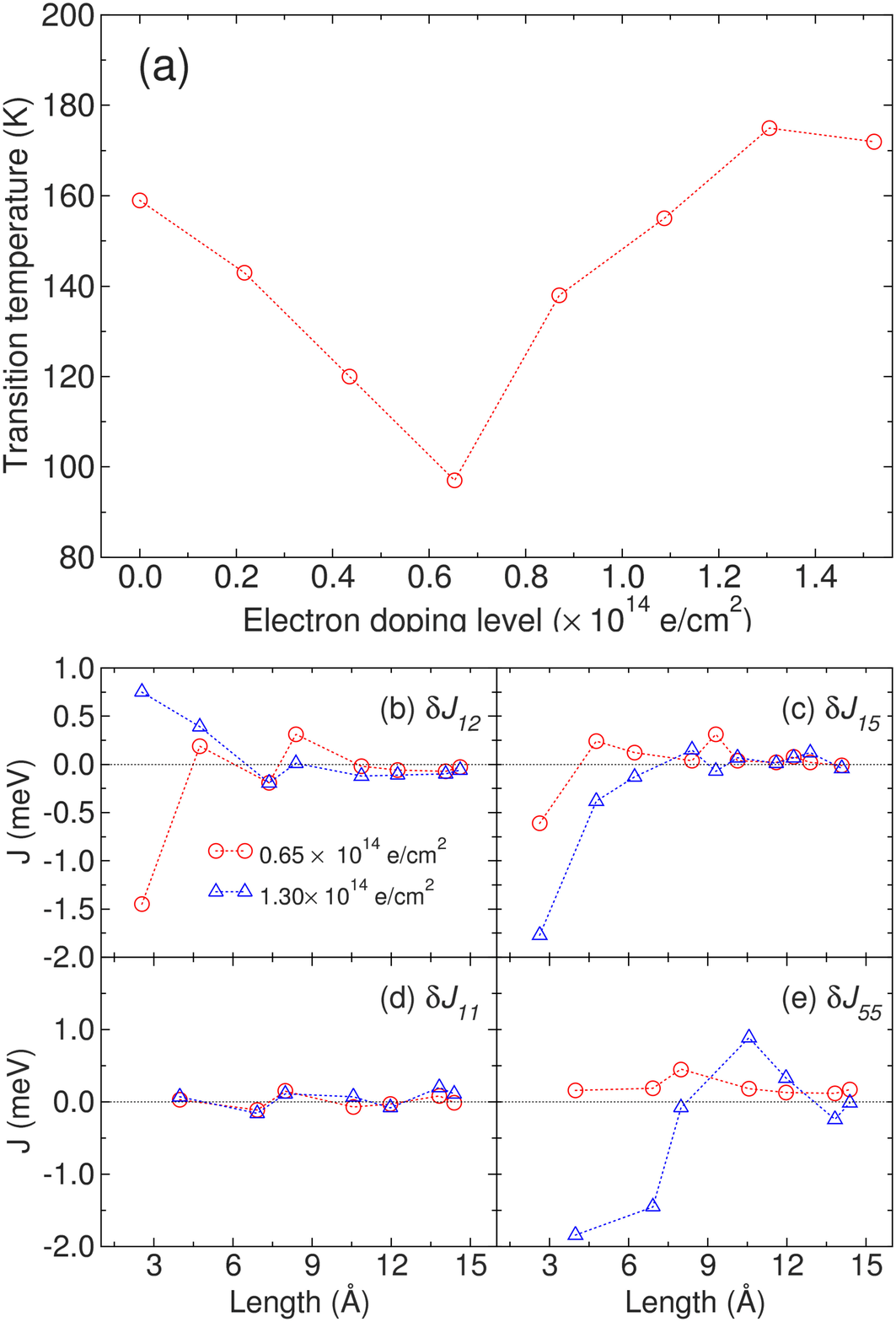}\\
  \caption{(Color online) (a) Transition temperature as a function of electron doping level.
  (b) $\delta J_{12}$, (c) $\delta J_{15}$, (d) $\delta J_{11}$, (e) $\delta J_{55}$ are the change of exchange interactions relative to the values at zero doping at electron doping $\delta$=0.65 and 1.30 $\times 10^{14} e/cm^{2}$.}
  \label{fig:layerdoping}
\end{figure}

{\it Electron doping:}
One of the most fascinating experiments in this material is that the Curie temperature can be
effectively tuned by electron doping. It has been demonstrated that the magnetic transition temperature
of a tri-layer FGT can be change from about 100 K to 300 K via electron doping\cite{deng2018gate}.
To understand this is of great importance both in fundamental science and potential applications.
We use the background electron approximate method to dope electron to a monolayer FGT, i.e., the surplus charges are
compensated by the positive uniform background charges.\cite{BlahaWIEN2k}
We calculate the exchange interactions under electron doping, and simulate the transition temperatures on a 36$\times$36 lattice by using re-MC method\cite{cao2009first}.
The transition temperature as a function of the electron doping level is shown in Fig.~\ref{fig:layerdoping}(a). The transition temperature goes down first and then goes up with the increasing of doping, in agreement with the experiment\cite{deng2018gate}.
Increasing electron doping levels  from 0 to 0.65 $\times$10$^{14}$ e/cm$^{2}$, the transition temperature decreases from
159 K to about 93 K.
Remarkably, further increase the electron doping level, the transition temperature increases rapidly from the lowest 93 K to about 175 K at 1.30$\times$10$^{14}$ e/cm$^{2}$ electron doping.

What causes the drastic change of Curie temperature by electron doping?
We first check the magnetic moments under electron doping.
As shown in Table~S3 in SM\cite{SM}, the magnetic moments of Fe1 and Fe2 ions are almost unchanged
under different doping level, whereas the magnetic moments of Fe5 ions  increases slightly and monotonically
with the increasing of electron doping level, from 1.53 $\mu_B$ at zero doping to 1.79 $\mu_B$ at about 1.30$\times$10$^{14}$ e/cm$^{2}$ doping.
The above results also suggest that the electron doping mainly affects the middle Fe5 plane (see Fig. 1).
However, the small and monotonic change of the magnetic moments could not account for the first decreasing then drastic increasing of Curie temperature.

We then look at the exchange interactions under the electron doping.
There are four non-equivalent of exchange interactions in the mono-layer FGT:
$J_{12}$, $J_{15}$, $J_{11}$ and $J_{55}$.
We show the change of these $J$s at $\delta$=0.65 $\times$10$^{14}$ e/cm$^{2}$ (red circles)
and 1.30$\times$10$^{14}$ e/cm$^{2}$ (blue triangles) relative to the values at zero doping in
Fig.~\ref{fig:layerdoping}(b-e),
which corresponding to the lowest Curie and highest Curie temperatures respectively.
To exam which interactions cause the drop of $T_c$ at
$\delta$=0.65 $\times$10$^{14}$ e/cm$^{2}$, we perform MC simulations by
using the values $J$s at this doping level one by one, while fix the values other $J$s
at zero doping.
The analyses show $J_{55}$ contribute most (about 29 K) to the decreasing of $T_c$,
whereas $J_{15}$ contribute about 14 K. These results are consistent with the
change of $J$s shown in Fig.~\ref{fig:layerdoping}, in which we see that all $J_{55}$ become larger
(i.e., less FM) at  $\delta$=0.65 $\times$10$^{14}$ e/cm$^{2}$.
For example, the FM interaction of the third neighboring $J_{55}$ reduces from
-0.92 meV to -0.46 meV, which contributes about 11 K to the decrease of $T_c$.

The most interesting result is that $T_c$ jump quickly from 93 K to
about 175 K from electron doping $\delta$=0.65$\times$10$^{14}$ e/cm$^{2}$ to $\delta$=1.30$\times$10$^{14}$ e/cm$^{2}$.
At this range of doping, we see the geometrically frustrated NN $J_{55}$ [see Fig.\ref{fig:layerdoping}(e)] reduce ``dramatically'' from about 2.5 meV to about 0.6 meV, which is nearly one fourth of its original  values, in accordance with the change of the transition temperatures.
Indeed, MC simulations show that when we only change the NN $J_{55}$ from about 2.5 meV to about 0.6 meV, while keep other exchange interactions at 
zero electron doping, the transition temperatures is about 176 K, and if we further
use the second NN $J_{55}$ at $\delta$=1.30$\times$10$^{14}$ e/cm$^{2}$, $T_c$ increase to 198 K.
This result is remarkable, that such small electron doping and changes of the magnetic interactions
can change the transition temperatures quite dramatically. We believe this effect should be general for the highly frustrated systems. It is possible to effectively tune the T$_c$ by changing the frustrated interactions via various techniques.

We note the Curie temperature can be tuned in a much larger temperature range, from around 100 K up to the room temperature for a tri-layer system\cite{deng2018gate}. As discussed in previous section, the inter-layer couplings in FGT are also frustrated. It is possible that the electron doping also suppresses the inter-layer frustrated coupling, and enhances the transition temperature. We leave this problem to future investigations.
It is also possible that the electron doping may change the MAE energies \cite{park2020controlling}. However, as shown in Fig.~S2 in SM\cite{SM}, the Curie temperature only change moderately with MAE energies.

\begin{figure}
  \centering
  \includegraphics[width=0.4\textwidth]{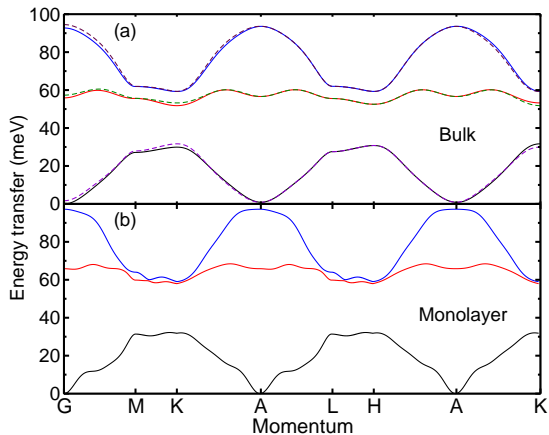}
  \caption{(Color online) The spin wave spectra of (a) bulk and (b) monolayer FGT.}\label{fig:spinwave}
\end{figure}

Using the calculated spin model parameters, we also calculate the magnon spectra of bulk and monolayer FGT on the basis of the Holstein-Primakoff transformation and the Fourier transformation. As shown in Fig. \ref{fig:spinwave}, there are three bands in magnetic excitation spectra of monolayer FGT, and the spin gap is quite small. The most remarkable feature is the near flat band around 63 meV.  With two monolayers in the unit cell, the bulk FGT has six spin-wave bands. Due to the smallness of interlayer magnetic coupling (i.e., $J_{23}$, $J_{56}$, etc.), the six bands in bulk FGT can be sorted as three groups, and each group bands has only very small splitting. Slightly modified the spin-wave dispersion, electronic doping do not change the near flat band in the magnon spectra. These results can be readily compared to future experiments.

To conclude, we carry out comprehensive investigation of both bulk and monolayer FGT to clarify several important issues about this fascinating materials by first-principles calculations. Both direct total energy calculations and linear response calculations confirm that the magnetic ground state of bulk FGT is ferromagnetic. The reason that previous calculations give AFM ground state is discussed. We unveil that there are frustrated magnetic interactions both within the single layer and between the layers in this compound. The electron doping mainly affect the middle Fe ion plane in a monolayer, and
tiny electron doping may tune the frustrated magnetic interactions resulting drastic change of the Curie temperatures.

This work was funded by the National Key Research and Development Program of China (Grants No. 2016YFB0201202), and
the Chinese National Science Foundation Grant number 11774327.
The numerical calculations have been done on the USTC HPC facilities. X.B. and X.W. were supported by the NSFC (Grants No. 11834006, No. 11525417, No.
51721001, and No. 11790311), National Key Research and Development Program of China (Grants No.
2018YFA0305704 and No. 2017YFA0303203). X.W. also acknowledges the support
from the Tencent Foundation through the XPLORER PRIZE.


\begin{thebibliography}{31}
\expandafter\ifx\csname natexlab\endcsname\relax\def\natexlab#1{#1}\fi
\expandafter\ifx\csname bibnamefont\endcsname\relax
  \def\bibnamefont#1{#1}\fi
\expandafter\ifx\csname bibfnamefont\endcsname\relax
  \def\bibfnamefont#1{#1}\fi
\expandafter\ifx\csname citenamefont\endcsname\relax
  \def\citenamefont#1{#1}\fi
\expandafter\ifx\csname url\endcsname\relax
  \def\url#1{\texttt{#1}}\fi
\expandafter\ifx\csname urlprefix\endcsname\relax\def\urlprefix{URL }\fi
\providecommand{\bibinfo}[2]{#2}
\providecommand{\eprint}[2][]{\url{#2}}

\bibitem[{\citenamefont{Gong et~al.}(2017)\citenamefont{Gong, Li, Li, Ji,
  Stern, Xia, Cao, Bao, Wang, Wang et~al.}}]{gong2017discovery}
\bibinfo{author}{\bibfnamefont{C.}~\bibnamefont{Gong}},
  \bibinfo{author}{\bibfnamefont{L.}~\bibnamefont{Li}},
  \bibinfo{author}{\bibfnamefont{Z.}~\bibnamefont{Li}},
  \bibinfo{author}{\bibfnamefont{H.}~\bibnamefont{Ji}},
  \bibinfo{author}{\bibfnamefont{A.}~\bibnamefont{Stern}},
  \bibinfo{author}{\bibfnamefont{Y.}~\bibnamefont{Xia}},
  \bibinfo{author}{\bibfnamefont{T.}~\bibnamefont{Cao}},
  \bibinfo{author}{\bibfnamefont{W.}~\bibnamefont{Bao}},
  \bibinfo{author}{\bibfnamefont{C.}~\bibnamefont{Wang}},
  \bibinfo{author}{\bibfnamefont{Y.}~\bibnamefont{Wang}}, \bibnamefont{et~al.},
  \bibinfo{journal}{Nature} \textbf{\bibinfo{volume}{546}},
  \bibinfo{pages}{265} (\bibinfo{year}{2017}).

\bibitem[{\citenamefont{Huang et~al.}(2017)\citenamefont{Huang, Clark,
  Navarromoratalla, Klein, Cheng, Seyler, Zhong, Schmidgall, Mcguire, Cobden
  et~al.}}]{huang2017layer-dependent}
\bibinfo{author}{\bibfnamefont{B.}~\bibnamefont{Huang}},
  \bibinfo{author}{\bibfnamefont{G.}~\bibnamefont{Clark}},
  \bibinfo{author}{\bibfnamefont{E.}~\bibnamefont{Navarromoratalla}},
  \bibinfo{author}{\bibfnamefont{D.~R.} \bibnamefont{Klein}},
  \bibinfo{author}{\bibfnamefont{R.}~\bibnamefont{Cheng}},
  \bibinfo{author}{\bibfnamefont{K.~L.} \bibnamefont{Seyler}},
  \bibinfo{author}{\bibfnamefont{D.}~\bibnamefont{Zhong}},
  \bibinfo{author}{\bibfnamefont{E.~R.} \bibnamefont{Schmidgall}},
  \bibinfo{author}{\bibfnamefont{M.~A.} \bibnamefont{Mcguire}},
  \bibinfo{author}{\bibfnamefont{D.}~\bibnamefont{Cobden}},
  \bibnamefont{et~al.}, \bibinfo{journal}{Nature}
  \textbf{\bibinfo{volume}{546}}, \bibinfo{pages}{270} (\bibinfo{year}{2017}).

\bibitem[{\citenamefont{Deng et~al.}(2018)\citenamefont{Deng, Yu, Song, Zhang,
  Wang, Sun, Yi, Wu, Wu, Zhu et~al.}}]{deng2018gate}
\bibinfo{author}{\bibfnamefont{Y.}~\bibnamefont{Deng}},
  \bibinfo{author}{\bibfnamefont{Y.}~\bibnamefont{Yu}},
  \bibinfo{author}{\bibfnamefont{Y.}~\bibnamefont{Song}},
  \bibinfo{author}{\bibfnamefont{J.}~\bibnamefont{Zhang}},
  \bibinfo{author}{\bibfnamefont{N.~Z.} \bibnamefont{Wang}},
  \bibinfo{author}{\bibfnamefont{Z.}~\bibnamefont{Sun}},
  \bibinfo{author}{\bibfnamefont{Y.}~\bibnamefont{Yi}},
  \bibinfo{author}{\bibfnamefont{Y.~Z.} \bibnamefont{Wu}},
  \bibinfo{author}{\bibfnamefont{S.}~\bibnamefont{Wu}},
  \bibinfo{author}{\bibfnamefont{J.}~\bibnamefont{Zhu}}, \bibnamefont{et~al.},
  \bibinfo{journal}{Nature} \textbf{\bibinfo{volume}{563}}, \bibinfo{pages}{94}
  (\bibinfo{year}{2018}).

\bibitem[{\citenamefont{Fei et~al.}(2018)\citenamefont{Fei, Huang, Malinowski,
  Wang, Song, Sanchez, Yao, Xiao, Zhu, May et~al.}}]{fei2018two}
\bibinfo{author}{\bibfnamefont{Z.}~\bibnamefont{Fei}},
  \bibinfo{author}{\bibfnamefont{B.}~\bibnamefont{Huang}},
  \bibinfo{author}{\bibfnamefont{P.}~\bibnamefont{Malinowski}},
  \bibinfo{author}{\bibfnamefont{W.}~\bibnamefont{Wang}},
  \bibinfo{author}{\bibfnamefont{T.}~\bibnamefont{Song}},
  \bibinfo{author}{\bibfnamefont{J.}~\bibnamefont{Sanchez}},
  \bibinfo{author}{\bibfnamefont{W.}~\bibnamefont{Yao}},
  \bibinfo{author}{\bibfnamefont{D.}~\bibnamefont{Xiao}},
  \bibinfo{author}{\bibfnamefont{X.}~\bibnamefont{Zhu}},
  \bibinfo{author}{\bibfnamefont{A.~F.} \bibnamefont{May}},
  \bibnamefont{et~al.}, \bibinfo{journal}{Nat. Mater.}
  \textbf{\bibinfo{volume}{17}}, \bibinfo{pages}{778} (\bibinfo{year}{2018}).

\bibitem[{\citenamefont{Kim et~al.}(2018)\citenamefont{Kim, Seo, Lee, Ko, Kim,
  Jang, Ok, Lee, Jo, Kang et~al.}}]{kim2018large}
\bibinfo{author}{\bibfnamefont{K.}~\bibnamefont{Kim}},
  \bibinfo{author}{\bibfnamefont{J.}~\bibnamefont{Seo}},
  \bibinfo{author}{\bibfnamefont{E.}~\bibnamefont{Lee}},
  \bibinfo{author}{\bibfnamefont{K.-T.} \bibnamefont{Ko}},
  \bibinfo{author}{\bibfnamefont{B.}~\bibnamefont{Kim}},
  \bibinfo{author}{\bibfnamefont{B.~G.} \bibnamefont{Jang}},
  \bibinfo{author}{\bibfnamefont{J.~M.} \bibnamefont{Ok}},
  \bibinfo{author}{\bibfnamefont{J.}~\bibnamefont{Lee}},
  \bibinfo{author}{\bibfnamefont{Y.~J.} \bibnamefont{Jo}},
  \bibinfo{author}{\bibfnamefont{W.}~\bibnamefont{Kang}}, \bibnamefont{et~al.},
  \bibinfo{journal}{Nat. Mater.} \textbf{\bibinfo{volume}{17}},
  \bibinfo{pages}{794} (\bibinfo{year}{2018}).

\bibitem[{\citenamefont{Gibertini et~al.}(2019)\citenamefont{Gibertini,
  Koperski, Morpurgo, and Novoselov}}]{gibertini2019magnetic}
\bibinfo{author}{\bibfnamefont{M.}~\bibnamefont{Gibertini}},
  \bibinfo{author}{\bibfnamefont{M.}~\bibnamefont{Koperski}},
  \bibinfo{author}{\bibfnamefont{A.~F.} \bibnamefont{Morpurgo}},
  \bibnamefont{and} \bibinfo{author}{\bibfnamefont{K.~S.}
  \bibnamefont{Novoselov}}, \bibinfo{journal}{Nat. Nanotechnol.}
  \textbf{\bibinfo{volume}{14}}, \bibinfo{pages}{408} (\bibinfo{year}{2019}).

\bibitem[{\citenamefont{Johansen et~al.}(2019)\citenamefont{Johansen,
  Risingg{\aa}rd, Sudb{\o}, Linder, and Brataas}}]{johansen2019current}
\bibinfo{author}{\bibfnamefont{{\O}.}~\bibnamefont{Johansen}},
  \bibinfo{author}{\bibfnamefont{V.}~\bibnamefont{Risingg{\aa}rd}},
  \bibinfo{author}{\bibfnamefont{A.}~\bibnamefont{Sudb{\o}}},
  \bibinfo{author}{\bibfnamefont{J.}~\bibnamefont{Linder}}, \bibnamefont{and}
  \bibinfo{author}{\bibfnamefont{A.}~\bibnamefont{Brataas}},
  \bibinfo{journal}{Phys. Rev. Lett.} \textbf{\bibinfo{volume}{122}},
  \bibinfo{pages}{217203} (\bibinfo{year}{2019}).

\bibitem[{\citenamefont{Ding et~al.}(2020)\citenamefont{Ding, Li, Xu, Li, Hou,
  Liu, Xi, Xu, Yao, and Wang}}]{ding2020observation}
\bibinfo{author}{\bibfnamefont{B.}~\bibnamefont{Ding}},
  \bibinfo{author}{\bibfnamefont{Z.}~\bibnamefont{Li}},
  \bibinfo{author}{\bibfnamefont{G.}~\bibnamefont{Xu}},
  \bibinfo{author}{\bibfnamefont{H.}~\bibnamefont{Li}},
  \bibinfo{author}{\bibfnamefont{Z.}~\bibnamefont{Hou}},
  \bibinfo{author}{\bibfnamefont{E.}~\bibnamefont{Liu}},
  \bibinfo{author}{\bibfnamefont{X.}~\bibnamefont{Xi}},
  \bibinfo{author}{\bibfnamefont{F.}~\bibnamefont{Xu}},
  \bibinfo{author}{\bibfnamefont{Y.}~\bibnamefont{Yao}}, \bibnamefont{and}
  \bibinfo{author}{\bibfnamefont{W.}~\bibnamefont{Wang}},
  \bibinfo{journal}{Nano Lett.} \textbf{\bibinfo{volume}{20}},
  \bibinfo{pages}{868} (\bibinfo{year}{2020}).

\bibitem[{\citenamefont{Fang et~al.}(2019)\citenamefont{Fang, Wan, Guo, Feng,
  Wang, Xing, Zhao, Dong, Yu, Zhao et~al.}}]{fang2019observation}
\bibinfo{author}{\bibfnamefont{C.}~\bibnamefont{Fang}},
  \bibinfo{author}{\bibfnamefont{C.}~\bibnamefont{Wan}},
  \bibinfo{author}{\bibfnamefont{C.}~\bibnamefont{Guo}},
  \bibinfo{author}{\bibfnamefont{C.}~\bibnamefont{Feng}},
  \bibinfo{author}{\bibfnamefont{X.}~\bibnamefont{Wang}},
  \bibinfo{author}{\bibfnamefont{Y.~W.} \bibnamefont{Xing}},
  \bibinfo{author}{\bibfnamefont{M.}~\bibnamefont{Zhao}},
  \bibinfo{author}{\bibfnamefont{J.}~\bibnamefont{Dong}},
  \bibinfo{author}{\bibfnamefont{G.}~\bibnamefont{Yu}},
  \bibinfo{author}{\bibfnamefont{Y.}~\bibnamefont{Zhao}}, \bibnamefont{et~al.},
  \bibinfo{journal}{Appl. Phys. Lett.} \textbf{\bibinfo{volume}{115}},
  \bibinfo{pages}{212402} (\bibinfo{year}{2019}).

\bibitem[{\citenamefont{Xu et~al.}(2019)\citenamefont{Xu, Phelan, and
  Chien}}]{xu2019large}
\bibinfo{author}{\bibfnamefont{J.}~\bibnamefont{Xu}},
  \bibinfo{author}{\bibfnamefont{W.~A.} \bibnamefont{Phelan}},
  \bibnamefont{and} \bibinfo{author}{\bibfnamefont{C.~L.} \bibnamefont{Chien}},
  \bibinfo{journal}{Nano Lett.} \textbf{\bibinfo{volume}{19}},
  \bibinfo{pages}{8250} (\bibinfo{year}{2019}).

\bibitem[{\citenamefont{Verchenko et~al.}(2015)\citenamefont{Verchenko,
  Tsirlin, Sobolev, Presniakov, and Shevelkov}}]{verchenko2015ferromagnetic}
\bibinfo{author}{\bibfnamefont{V.~Y.} \bibnamefont{Verchenko}},
  \bibinfo{author}{\bibfnamefont{A.~A.} \bibnamefont{Tsirlin}},
  \bibinfo{author}{\bibfnamefont{A.~V.} \bibnamefont{Sobolev}},
  \bibinfo{author}{\bibfnamefont{I.~A.} \bibnamefont{Presniakov}},
  \bibnamefont{and} \bibinfo{author}{\bibfnamefont{A.~V.}
  \bibnamefont{Shevelkov}}, \bibinfo{journal}{Inorg. Chem.}
  \textbf{\bibinfo{volume}{54}}, \bibinfo{pages}{8598} (\bibinfo{year}{2015}).

\bibitem[{\citenamefont{Park et~al.}(2020)\citenamefont{Park, Kim, Liu, Hwang,
  Kim, Kim, Kim, Petrovic, Hwang, Mo et~al.}}]{park2020controlling}
\bibinfo{author}{\bibfnamefont{S.~Y.} \bibnamefont{Park}},
  \bibinfo{author}{\bibfnamefont{D.~S.} \bibnamefont{Kim}},
  \bibinfo{author}{\bibfnamefont{Y.}~\bibnamefont{Liu}},
  \bibinfo{author}{\bibfnamefont{J.}~\bibnamefont{Hwang}},
  \bibinfo{author}{\bibfnamefont{Y.}~\bibnamefont{Kim}},
  \bibinfo{author}{\bibfnamefont{W.}~\bibnamefont{Kim}},
  \bibinfo{author}{\bibfnamefont{J.}~\bibnamefont{Kim}},
  \bibinfo{author}{\bibfnamefont{C.}~\bibnamefont{Petrovic}},
  \bibinfo{author}{\bibfnamefont{C.}~\bibnamefont{Hwang}},
  \bibinfo{author}{\bibfnamefont{S.~K.} \bibnamefont{Mo}},
  \bibnamefont{et~al.}, \bibinfo{journal}{Nano Lett.}
  \textbf{\bibinfo{volume}{20}}, \bibinfo{pages}{95} (\bibinfo{year}{2020}).

\bibitem[{\citenamefont{Zhu et~al.}(2016)\citenamefont{Zhu, Janoschek, Chaves,
  Cezar, Durakiewicz, Ronning, Sassa, Mansson, Scott, Wakeham
  et~al.}}]{zhu2016electronic}
\bibinfo{author}{\bibfnamefont{J.-X.} \bibnamefont{Zhu}},
  \bibinfo{author}{\bibfnamefont{M.}~\bibnamefont{Janoschek}},
  \bibinfo{author}{\bibfnamefont{D.~S.} \bibnamefont{Chaves}},
  \bibinfo{author}{\bibfnamefont{J.~C.} \bibnamefont{Cezar}},
  \bibinfo{author}{\bibfnamefont{T.}~\bibnamefont{Durakiewicz}},
  \bibinfo{author}{\bibfnamefont{F.}~\bibnamefont{Ronning}},
  \bibinfo{author}{\bibfnamefont{Y.}~\bibnamefont{Sassa}},
  \bibinfo{author}{\bibfnamefont{M.}~\bibnamefont{Mansson}},
  \bibinfo{author}{\bibfnamefont{B.~L.} \bibnamefont{Scott}},
  \bibinfo{author}{\bibfnamefont{N.}~\bibnamefont{Wakeham}},
  \bibnamefont{et~al.}, \bibinfo{journal}{Phys. Rev. B}
  \textbf{\bibinfo{volume}{93}}, \bibinfo{pages}{144404}
  (\bibinfo{year}{2016}).

\bibitem[{\citenamefont{Zhuang et~al.}(2016)\citenamefont{Zhuang, Kent, and
  Hennig}}]{zhuang2016strong}
\bibinfo{author}{\bibfnamefont{H.~L.} \bibnamefont{Zhuang}},
  \bibinfo{author}{\bibfnamefont{P.~R.~C.} \bibnamefont{Kent}},
  \bibnamefont{and} \bibinfo{author}{\bibfnamefont{R.~G.}
  \bibnamefont{Hennig}}, \bibinfo{journal}{Phys. Rev. B}
  \textbf{\bibinfo{volume}{93}}, \bibinfo{pages}{134407}
  (\bibinfo{year}{2016}).

\bibitem[{\citenamefont{Liu et~al.}(2017{\natexlab{a}})\citenamefont{Liu, Zou,
  Zhou, Zhang, Wang, Li, Qu, and Zhang}}]{liu2017critical}
\bibinfo{author}{\bibfnamefont{B.}~\bibnamefont{Liu}},
  \bibinfo{author}{\bibfnamefont{Y.}~\bibnamefont{Zou}},
  \bibinfo{author}{\bibfnamefont{S.}~\bibnamefont{Zhou}},
  \bibinfo{author}{\bibfnamefont{L.}~\bibnamefont{Zhang}},
  \bibinfo{author}{\bibfnamefont{Z.}~\bibnamefont{Wang}},
  \bibinfo{author}{\bibfnamefont{H.}~\bibnamefont{Li}},
  \bibinfo{author}{\bibfnamefont{Z.}~\bibnamefont{Qu}}, \bibnamefont{and}
  \bibinfo{author}{\bibfnamefont{Y.}~\bibnamefont{Zhang}},
  \bibinfo{journal}{Sci Rep} \textbf{\bibinfo{volume}{7}},
  \bibinfo{pages}{6184} (\bibinfo{year}{2017}{\natexlab{a}}).

\bibitem[{\citenamefont{Liu et~al.}(2017{\natexlab{b}})\citenamefont{Liu, Yuan,
  Zou, Sheng, Huang, Zhang, Ling, Liu, Wang, Zhang et~al.}}]{liu2017wafer}
\bibinfo{author}{\bibfnamefont{S.}~\bibnamefont{Liu}},
  \bibinfo{author}{\bibfnamefont{X.}~\bibnamefont{Yuan}},
  \bibinfo{author}{\bibfnamefont{Y.}~\bibnamefont{Zou}},
  \bibinfo{author}{\bibfnamefont{Y.}~\bibnamefont{Sheng}},
  \bibinfo{author}{\bibfnamefont{C.}~\bibnamefont{Huang}},
  \bibinfo{author}{\bibfnamefont{E.}~\bibnamefont{Zhang}},
  \bibinfo{author}{\bibfnamefont{J.}~\bibnamefont{Ling}},
  \bibinfo{author}{\bibfnamefont{Y.}~\bibnamefont{Liu}},
  \bibinfo{author}{\bibfnamefont{W.}~\bibnamefont{Wang}},
  \bibinfo{author}{\bibfnamefont{C.}~\bibnamefont{Zhang}},
  \bibnamefont{et~al.}, \bibinfo{journal}{npj 2D Mater. Appl.}
  \textbf{\bibinfo{volume}{1}}, \bibinfo{pages}{30}
  (\bibinfo{year}{2017}{\natexlab{b}}).

\bibitem[{\citenamefont{Yi et~al.}(2017)\citenamefont{Yi, Zhuang, Zou, Wu, Cao,
  Tang, Calder, Kent, Mandrus, and Gai}}]{yi2016competing}
\bibinfo{author}{\bibfnamefont{J.}~\bibnamefont{Yi}},
  \bibinfo{author}{\bibfnamefont{H.}~\bibnamefont{Zhuang}},
  \bibinfo{author}{\bibfnamefont{Q.}~\bibnamefont{Zou}},
  \bibinfo{author}{\bibfnamefont{Z.}~\bibnamefont{Wu}},
  \bibinfo{author}{\bibfnamefont{G.}~\bibnamefont{Cao}},
  \bibinfo{author}{\bibfnamefont{S.}~\bibnamefont{Tang}},
  \bibinfo{author}{\bibfnamefont{S.}~\bibnamefont{Calder}},
  \bibinfo{author}{\bibfnamefont{P.}~\bibnamefont{Kent}},
  \bibinfo{author}{\bibfnamefont{D.}~\bibnamefont{Mandrus}}, \bibnamefont{and}
  \bibinfo{author}{\bibfnamefont{Z.}~\bibnamefont{Gai}}, \bibinfo{journal}{2D
  Mater.} \textbf{\bibinfo{volume}{4}}, \bibinfo{pages}{011005}
  (\bibinfo{year}{2017}).

\bibitem[{\citenamefont{Tan et~al.}(2018)\citenamefont{Tan, Lee, Jung, Park,
  Albarakati, Partridge, Field, McCulloch, Wang, and Lee}}]{tan2018hard}
\bibinfo{author}{\bibfnamefont{C.}~\bibnamefont{Tan}},
  \bibinfo{author}{\bibfnamefont{J.}~\bibnamefont{Lee}},
  \bibinfo{author}{\bibfnamefont{S.-G.} \bibnamefont{Jung}},
  \bibinfo{author}{\bibfnamefont{T.}~\bibnamefont{Park}},
  \bibinfo{author}{\bibfnamefont{S.}~\bibnamefont{Albarakati}},
  \bibinfo{author}{\bibfnamefont{J.}~\bibnamefont{Partridge}},
  \bibinfo{author}{\bibfnamefont{M.~R.} \bibnamefont{Field}},
  \bibinfo{author}{\bibfnamefont{D.~G.} \bibnamefont{McCulloch}},
  \bibinfo{author}{\bibfnamefont{L.}~\bibnamefont{Wang}}, \bibnamefont{and}
  \bibinfo{author}{\bibfnamefont{C.}~\bibnamefont{Lee}}, \bibinfo{journal}{Nat.
  Commun.} \textbf{\bibinfo{volume}{9}}, \bibinfo{pages}{1554}
  (\bibinfo{year}{2018}).

\bibitem[{\citenamefont{Zhang et~al.}(2018)\citenamefont{Zhang, Lu, Zhu, Tan,
  Feng, Liu, Zhang, Chen, Liu, Luo et~al.}}]{zhang2018emergence}
\bibinfo{author}{\bibfnamefont{Y.}~\bibnamefont{Zhang}},
  \bibinfo{author}{\bibfnamefont{H.}~\bibnamefont{Lu}},
  \bibinfo{author}{\bibfnamefont{X.}~\bibnamefont{Zhu}},
  \bibinfo{author}{\bibfnamefont{S.}~\bibnamefont{Tan}},
  \bibinfo{author}{\bibfnamefont{W.}~\bibnamefont{Feng}},
  \bibinfo{author}{\bibfnamefont{Q.}~\bibnamefont{Liu}},
  \bibinfo{author}{\bibfnamefont{W.}~\bibnamefont{Zhang}},
  \bibinfo{author}{\bibfnamefont{Q.}~\bibnamefont{Chen}},
  \bibinfo{author}{\bibfnamefont{Y.}~\bibnamefont{Liu}},
  \bibinfo{author}{\bibfnamefont{X.}~\bibnamefont{Luo}}, \bibnamefont{et~al.},
  \bibinfo{journal}{Sci. Adv.} \textbf{\bibinfo{volume}{4}},
  \bibinfo{pages}{eaao6791} (\bibinfo{year}{2018}).

\bibitem[{\citenamefont{Albarakati et~al.}(2019)\citenamefont{Albarakati, Tan,
  Chen, Partridge, Zheng, Farrar, Mayes, Field, Lee, Wang
  et~al.}}]{albarakati2019antisymmetric}
\bibinfo{author}{\bibfnamefont{S.}~\bibnamefont{Albarakati}},
  \bibinfo{author}{\bibfnamefont{C.}~\bibnamefont{Tan}},
  \bibinfo{author}{\bibfnamefont{Z.-J.} \bibnamefont{Chen}},
  \bibinfo{author}{\bibfnamefont{J.~G.} \bibnamefont{Partridge}},
  \bibinfo{author}{\bibfnamefont{G.}~\bibnamefont{Zheng}},
  \bibinfo{author}{\bibfnamefont{L.}~\bibnamefont{Farrar}},
  \bibinfo{author}{\bibfnamefont{E.~L.} \bibnamefont{Mayes}},
  \bibinfo{author}{\bibfnamefont{M.~R.} \bibnamefont{Field}},
  \bibinfo{author}{\bibfnamefont{C.}~\bibnamefont{Lee}},
  \bibinfo{author}{\bibfnamefont{Y.}~\bibnamefont{Wang}}, \bibnamefont{et~al.},
  \bibinfo{journal}{Sci. Adv.} \textbf{\bibinfo{volume}{5}},
  \bibinfo{pages}{eaaw0409} (\bibinfo{year}{2019}).

\bibitem[{\citenamefont{Alghamdi et~al.}(2019)\citenamefont{Alghamdi, Lohmann,
  Li, Jothi, Shao, Aldosary, Su, Fokwa, and Shi}}]{alghamdi2019highly}
\bibinfo{author}{\bibfnamefont{M.}~\bibnamefont{Alghamdi}},
  \bibinfo{author}{\bibfnamefont{M.}~\bibnamefont{Lohmann}},
  \bibinfo{author}{\bibfnamefont{J.}~\bibnamefont{Li}},
  \bibinfo{author}{\bibfnamefont{P.~R.} \bibnamefont{Jothi}},
  \bibinfo{author}{\bibfnamefont{Q.}~\bibnamefont{Shao}},
  \bibinfo{author}{\bibfnamefont{M.}~\bibnamefont{Aldosary}},
  \bibinfo{author}{\bibfnamefont{T.}~\bibnamefont{Su}},
  \bibinfo{author}{\bibfnamefont{B.~P.~T.} \bibnamefont{Fokwa}},
  \bibnamefont{and} \bibinfo{author}{\bibfnamefont{J.}~\bibnamefont{Shi}},
  \bibinfo{journal}{Nano Lett.} \textbf{\bibinfo{volume}{19}},
  \bibinfo{pages}{4400} (\bibinfo{year}{2019}).

\bibitem[{\citenamefont{Chen et~al.}(2013)\citenamefont{Chen, Yang, Wang, Imai,
  Ohta, Michioka, Yoshimura, and Fang}}]{chen2013magnetic}
\bibinfo{author}{\bibfnamefont{B.}~\bibnamefont{Chen}},
  \bibinfo{author}{\bibfnamefont{J.}~\bibnamefont{Yang}},
  \bibinfo{author}{\bibfnamefont{H.}~\bibnamefont{Wang}},
  \bibinfo{author}{\bibfnamefont{M.}~\bibnamefont{Imai}},
  \bibinfo{author}{\bibfnamefont{H.}~\bibnamefont{Ohta}},
  \bibinfo{author}{\bibfnamefont{C.}~\bibnamefont{Michioka}},
  \bibinfo{author}{\bibfnamefont{K.}~\bibnamefont{Yoshimura}},
  \bibnamefont{and} \bibinfo{author}{\bibfnamefont{M.}~\bibnamefont{Fang}},
  \bibinfo{journal}{J. Phys. Soc. Jpn.} \textbf{\bibinfo{volume}{82}},
  \bibinfo{pages}{124711} (\bibinfo{year}{2013}).

\bibitem[{\citenamefont{Wang et~al.}(2019)\citenamefont{Wang, Tang, Xia, He,
  Zhang, Liu, Wan, Fang, Guo, Yang et~al.}}]{wang2019current-driven}
\bibinfo{author}{\bibfnamefont{X.}~\bibnamefont{Wang}},
  \bibinfo{author}{\bibfnamefont{J.}~\bibnamefont{Tang}},
  \bibinfo{author}{\bibfnamefont{X.}~\bibnamefont{Xia}},
  \bibinfo{author}{\bibfnamefont{C.}~\bibnamefont{He}},
  \bibinfo{author}{\bibfnamefont{J.}~\bibnamefont{Zhang}},
  \bibinfo{author}{\bibfnamefont{Y.}~\bibnamefont{Liu}},
  \bibinfo{author}{\bibfnamefont{C.}~\bibnamefont{Wan}},
  \bibinfo{author}{\bibfnamefont{C.}~\bibnamefont{Fang}},
  \bibinfo{author}{\bibfnamefont{C.}~\bibnamefont{Guo}},
  \bibinfo{author}{\bibfnamefont{W.}~\bibnamefont{Yang}}, \bibnamefont{et~al.},
  \bibinfo{journal}{Sci. Adv.} \textbf{\bibinfo{volume}{5}},
  \bibinfo{pages}{eaaw8904} (\bibinfo{year}{2019}).

\bibitem[{\citenamefont{Le{\'o}n-Brito
  et~al.}(2016)\citenamefont{Le{\'o}n-Brito, Bauer, Ronning, Thompson, and
  Movshovich}}]{leon2016magnetic}
\bibinfo{author}{\bibfnamefont{N.}~\bibnamefont{Le{\'o}n-Brito}},
  \bibinfo{author}{\bibfnamefont{E.~D.} \bibnamefont{Bauer}},
  \bibinfo{author}{\bibfnamefont{F.}~\bibnamefont{Ronning}},
  \bibinfo{author}{\bibfnamefont{J.~D.} \bibnamefont{Thompson}},
  \bibnamefont{and}
  \bibinfo{author}{\bibfnamefont{R.}~\bibnamefont{Movshovich}},
  \bibinfo{journal}{J. Appl. Phys.} \textbf{\bibinfo{volume}{120}},
  \bibinfo{pages}{083903} (\bibinfo{year}{2016}).

\bibitem[{\citenamefont{Jang et~al.}(2020)\citenamefont{Jang, Yoon, Jeong,
  Ryee, Kim, and Han}}]{C9NR10171C}
\bibinfo{author}{\bibfnamefont{S.~W.} \bibnamefont{Jang}},
  \bibinfo{author}{\bibfnamefont{H.}~\bibnamefont{Yoon}},
  \bibinfo{author}{\bibfnamefont{M.~Y.} \bibnamefont{Jeong}},
  \bibinfo{author}{\bibfnamefont{S.}~\bibnamefont{Ryee}},
  \bibinfo{author}{\bibfnamefont{H.-S.} \bibnamefont{Kim}}, \bibnamefont{and}
  \bibinfo{author}{\bibfnamefont{M.~J.} \bibnamefont{Han}},
  \bibinfo{journal}{Nanoscale} \textbf{\bibinfo{volume}{12}},
  \bibinfo{pages}{13501} (\bibinfo{year}{2020}).

\bibitem[{\citenamefont{Deiseroth et~al.}(2006)\citenamefont{Deiseroth,
  Aleksandrov, Reiner, Kienle, and Kremer}}]{deiseroth2006fe3gete2}
\bibinfo{author}{\bibfnamefont{H.-J.} \bibnamefont{Deiseroth}},
  \bibinfo{author}{\bibfnamefont{K.}~\bibnamefont{Aleksandrov}},
  \bibinfo{author}{\bibfnamefont{C.}~\bibnamefont{Reiner}},
  \bibinfo{author}{\bibfnamefont{L.}~\bibnamefont{Kienle}}, \bibnamefont{and}
  \bibinfo{author}{\bibfnamefont{R.~K.} \bibnamefont{Kremer}},
  \bibinfo{journal}{Eur. J. Inorg. Chem.} \textbf{\bibinfo{volume}{2006}},
  \bibinfo{pages}{1561} (\bibinfo{year}{2006}).

\bibitem[{SM()}]{SM}
\bibinfo{note}{In the supplement materail, we present computational details and
  some additional results, including the exchange inetractions ect.}

\bibitem[{\citenamefont{Wan et~al.}(2006)\citenamefont{Wan, Yin, and
  Savrasov}}]{wan2006calculation}
\bibinfo{author}{\bibfnamefont{X.}~\bibnamefont{Wan}},
  \bibinfo{author}{\bibfnamefont{Q.}~\bibnamefont{Yin}}, \bibnamefont{and}
  \bibinfo{author}{\bibfnamefont{S.~Y.} \bibnamefont{Savrasov}},
  \bibinfo{journal}{Phys. Rev. Lett.} \textbf{\bibinfo{volume}{97}},
  \bibinfo{pages}{266403} (\bibinfo{year}{2006}).

\bibitem[{\citenamefont{Cao et~al.}(2009)\citenamefont{Cao, Guo, Vanderbilt,
  and He}}]{cao2009first}
\bibinfo{author}{\bibfnamefont{K.}~\bibnamefont{Cao}},
  \bibinfo{author}{\bibfnamefont{G.-C.} \bibnamefont{Guo}},
  \bibinfo{author}{\bibfnamefont{D.}~\bibnamefont{Vanderbilt}},
  \bibnamefont{and} \bibinfo{author}{\bibfnamefont{L.}~\bibnamefont{He}},
  \bibinfo{journal}{Phys. Rev. Lett.} \textbf{\bibinfo{volume}{103}},
  \bibinfo{pages}{257201} (\bibinfo{year}{2009}).

\bibitem[{\citenamefont{Yang et~al.}(2001)\citenamefont{Yang, Savrasov, and
  Kotliar}}]{MAE-U}
\bibinfo{author}{\bibfnamefont{I.}~\bibnamefont{Yang}},
  \bibinfo{author}{\bibfnamefont{S.~Y.} \bibnamefont{Savrasov}},
  \bibnamefont{and} \bibinfo{author}{\bibfnamefont{G.}~\bibnamefont{Kotliar}},
  \bibinfo{journal}{Phys. Rev. Lett.} \textbf{\bibinfo{volume}{87}},
  \bibinfo{pages}{216405} (\bibinfo{year}{2001}).

\bibitem[{\citenamefont{Blaha et~al.}(2001)\citenamefont{Blaha, Schwarz,
  Madsen, Kvasnicka, and Luitz}}]{BlahaWIEN2k}
\bibinfo{author}{\bibfnamefont{P.}~\bibnamefont{Blaha}},
  \bibinfo{author}{\bibfnamefont{K.}~\bibnamefont{Schwarz}},
  \bibinfo{author}{\bibfnamefont{G.}~\bibnamefont{Madsen}},
  \bibinfo{author}{\bibfnamefont{D.}~\bibnamefont{Kvasnicka}},
  \bibnamefont{and} \bibinfo{author}{\bibfnamefont{J.}~\bibnamefont{Luitz}},
  \bibinfo{journal}{Tech. Universitat Wien, Austria}
  \textbf{\bibinfo{volume}{28}} (\bibinfo{year}{2001}).

\end{thebibliography}

\end{document}